\documentclass[]{spie}  
\usepackage[]{graphicx}
\usepackage{color}

\title{NIRISS aperture masking interferometry: an overview of science opportunities} 

\author{\'Etienne Artigau\supit{a}, Anand Sivaramakrishnan\supit{b}, Alexandra Z. Greenbaum\supit{c}, Ren\'e Doyon\supit{a}, Paul Goudfrooij\supit{b}, Alex W. Fullerton\supit{b}, David Lafreni\`ere\supit{a}, Kevin Volk\supit{b}, Lo\"{i}c Albert\supit{a}, Andr\'e Martel\supit{b},  K. E. Saavik Ford\supit{d} \& Barry L. McKernan\supit{d}
\skiplinehalf
\supit{a}D\'epartement de Physique, Universit\'e de Montr\'eal, Montr\'eal H3C 3J7, Canada; \\
\supit{b}Space Telescope Science Institute, 3700 San Martin Drive, Baltimore, MD 21218, USA;\\
\supit{c}Johns Hopkins University Department of Physics and Astronomy 3400 N. Charles, Baltimore, MD 21218;\\
\supit{d}Dept. of Science, Borough of Manhattan Community College, City University of New York, 199 Chambers Street, NY 10007
}

\authorinfo{Further author information: (Send correspondence to E.A., artigau@astro.umontreal.ca, Telephone: 1 514-343-6111, ext. 3190}
 
\begin{document} 
\maketitle 

\begin{abstract}

JWST's Near-Infrared Imager and Slitless Spectrograph (NIRISS) includes an Aperture Masking Interferometry (AMI) mode designed to be used between 2.7$\mu$m and 4.8$\mu$m. At these wavelengths,  it will have the highest angular resolution of any mode on JWST, and, for faint targets, of any  existing or planned infrastructure. NIRISS AMI is uniquely suited to detect thermal emission of young massive planets and will permit the characterization of the mid-IR flux of exoplanets discovered by the GPI and SPHERE adaptive optics surveys. It will also directly detect massive planets found by GAIA through  astrometric accelerations, providing the first opportunity ever to  get both a mass and a flux measurement for non-transiting giant planets. NIRISS AMI will also enable the study of the nuclear environment of AGNs.
\end{abstract}


\section{INTRODUCTION}
\label{sec:intro}  
\begin{figure}
\begin{center}
\begin{tabular}{c}
\includegraphics[height=7cm]{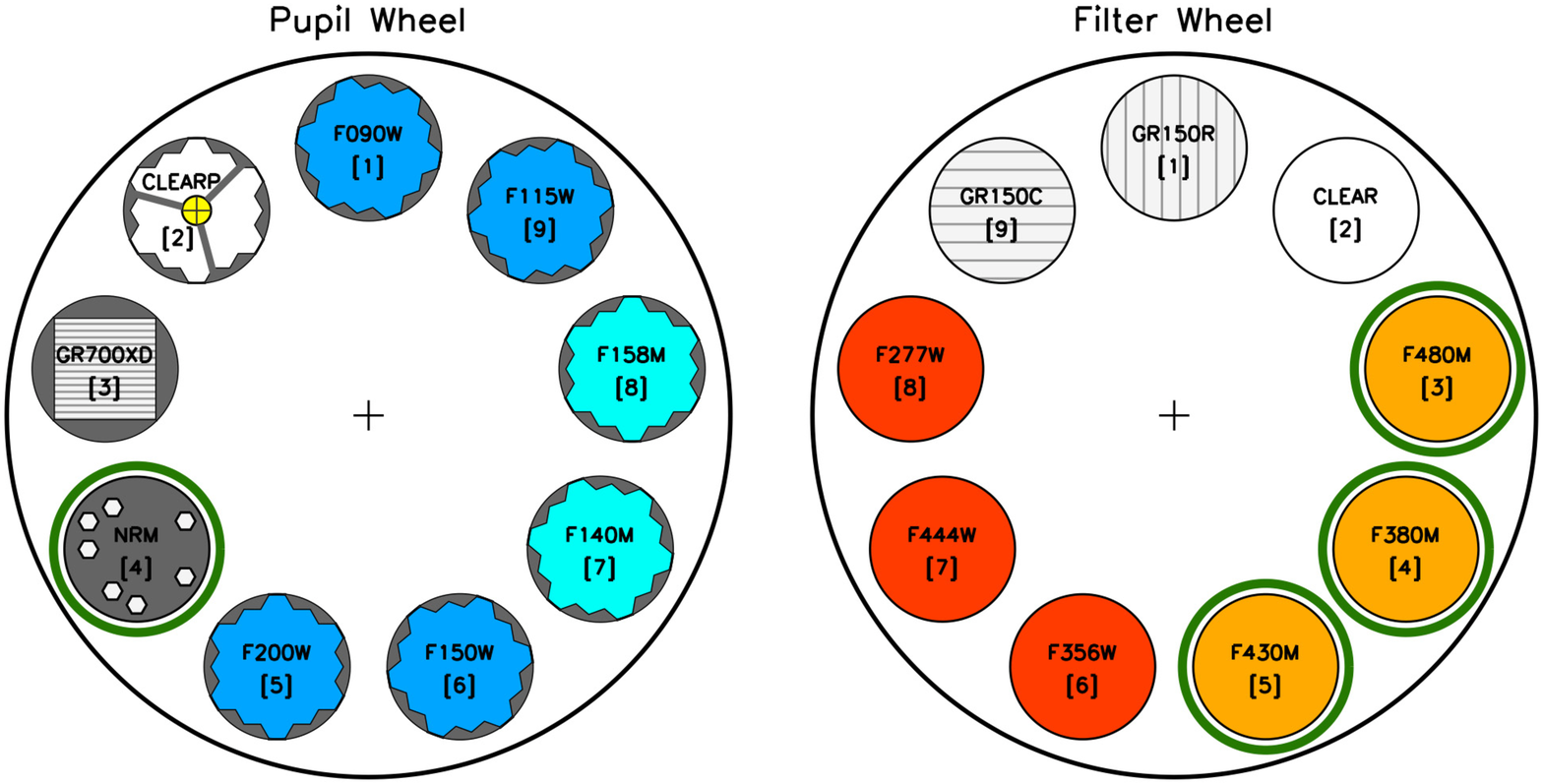}\\
\includegraphics[width=15cm]{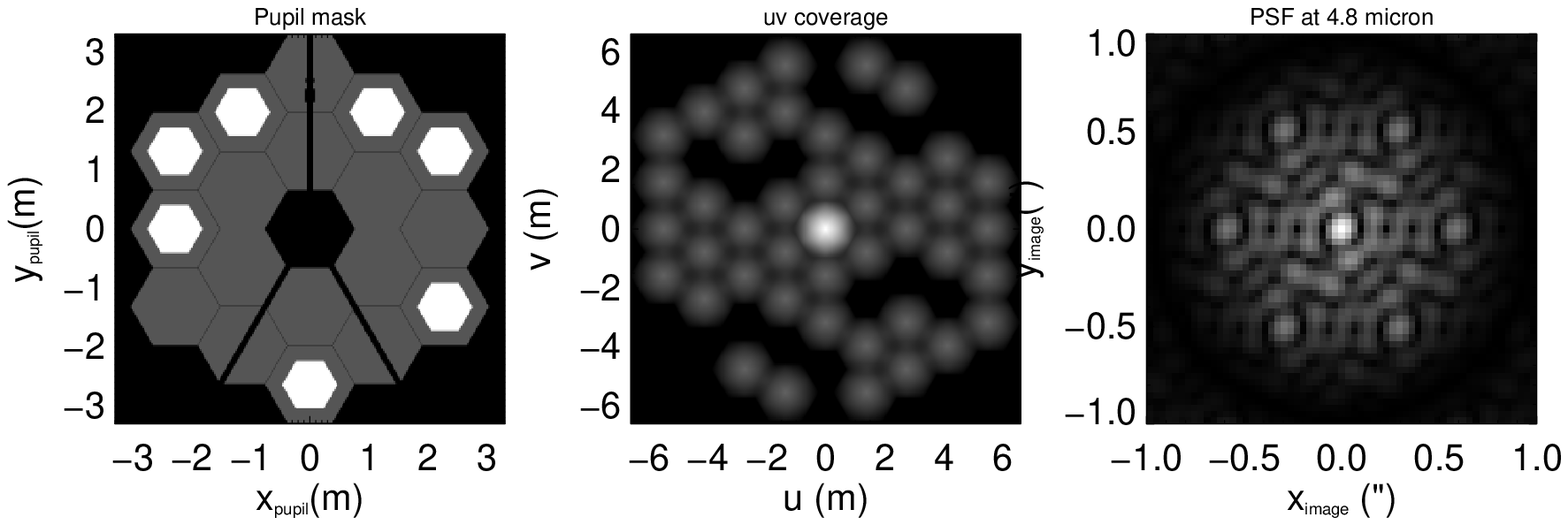}
\end{tabular}
\end{center}
\caption[example]{ \label{fig1} {\it Top:}  Schematics of the NIRISS wheels. The  {\it pupil wheel} (left) contains photometric filters blueward of $2.5\mu$m, that can be combined with the low-resolution grisms (GR150 on the filter wheel), the intermediate resolution grism GR700 for single-object slit-less spectroscopy (SOSS), an undersized pupil for use in imaging with redder filters and, lastly, the NRM mask with its 7 apertures. The {\it filter wheel}  includes two low-resolution grisms (WFSS) and six filters redward of $2.5\mu$m. The NRM mask is designed to be used with 3 of these filters (F380M, F430M, F480M) providing non-redundant baseline through the bandpass. {\it Bottom left:} NIRISS NRM mask projected on the JWST pupil (white hexagons). The outline of the full JWST pupil is also shown (dark grey). {\it Bottom center:} $u-v$ coverage of the JWST pupil, the pupil coverage is non-redundant. {\it Bottom right:} Oversampled NIRISS AMI point-spread function.}
\end{figure}

The Near-Infrared Imager and Slitless Spectrograph (NIRISS\cite{Doyon:2012}) is one of the four instruments on board of the James Webb Space Telescope. NIRISS includes a near-infrared ($1-2.5\mu$m) Wide-Field Slitless Spectroscopy (WFSS) mode at low resolution, mostly tuned for the detection of first-light galaxy and the study of galaxy assembly. There is a defocussed Single-Object Slitless Spectroscopy (SOSS) mode for high-accuracy transit spectroscopy, largely targeted toward transit spectroscopy of bright exoplanet host stars. The third mode is the non-redundant mask (NRM) aperture masking interferometry (AMI). By masking most of the pupil except certain positions, this masks lifts the ambiguity between pupil-space aberrations and image-plane point-spread function thus side-stepping the limitations imposed by speckles in filled-pupil observations (see figure~\ref{fig1}, bottom). In this mode, most of the pupil is masked, leaving open a set of 7 holes that spawn non-redundant baselines across the pupil, sampling 21 baselines with sizes projected on the JWST pupil ranging from 1.32\,m to 5.28\,m  (center to center). The 7 hexagonal holes together  cover 15\% of the pupil surface and each  have a projected flat-to-flat width of 0.8\,m.  The AMI is tuned to detect faint structures of simple geometry between $0.5$ and a few $\lambda/D$ next to bright sources, either planetary companions to stellar hosts\cite{Sivaramakrishnan:2010} or bars and disks within active galaxy nuclei\cite{Ford:2014}. AMI will also observe the immediate environment of supermassive black holes (SMBHs) in brighter active galactic nuclei (AGNs\cite{Ford:2014}), which should afford tests of cosmological theories and insights into galaxy evolution. Envisioned exoplanet work with the NIRISS AMI includes the mid-infrared follow-up of exoplanets uncovered by the ongoing campaigns with the GPI\cite{Macintosh:2014} (Gemini) and SPHERE\cite{Beuzit:2008, Beuzit:2010} (ESO-VLT) high-contrast imagers, as well as direct detection of planets uncovered by astrometric acceleration by the GAIA mission.

The NIRISS AMI mask is designed to be used in conjunction with 3 medium-band filters at $3.8\mu$m, $4.3\mu$m and $4.8\mu$m, but can also be used with degraded sensitivity with a broad band filter centered at $2.77\mu$m\cite{Greenbaum:2014} (see Figure~\ref{fig1} for the NIRISS pupil and filter wheel setup). In addition to its science usage, the AMI mode in NIRISS is also foreseen as a segment-phasing option in addition to the nominal JWST segment phasing strategy\cite{Sivaramakrishnan:2012}, thus reducing mission-level risks.


\section{NIRISS NRM filter choices}
\begin{figure}[!h]
\begin{center}
\begin{tabular}{c}
\includegraphics[height=8cm]{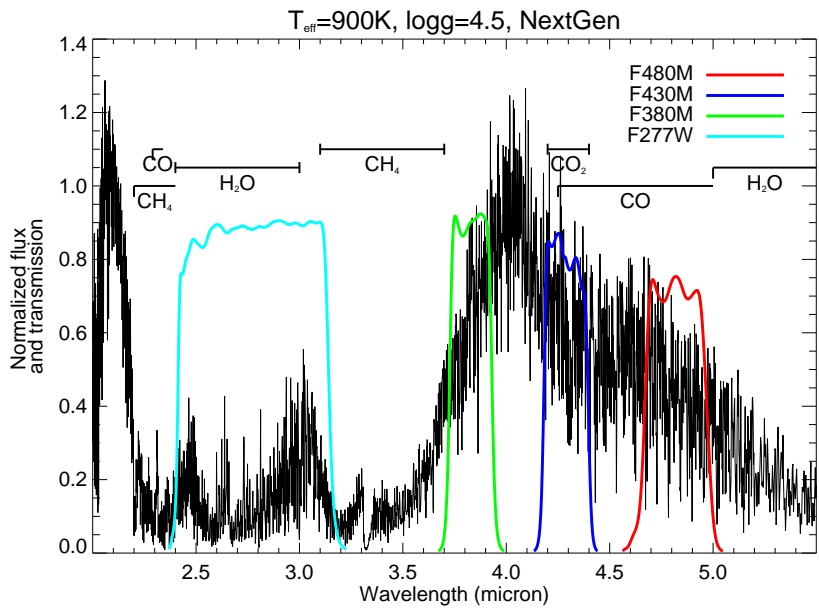}
\end{tabular}
\end{center}
\caption{ \label{fig3}  Theoretical spectrum\cite{Allard:2011} of an ultracool dwarf with an effective temperature of 900\,K and a surface gravity of $\log$ g=4.5 between 2.0 and 5.5$\mu$m overplotted with the bandpasses of filters that can be used with the NIRISS AMI mode.}
\end{figure}

\begin{figure}[!h]
\begin{center}
\begin{tabular}{c}
\includegraphics[height=8cm]{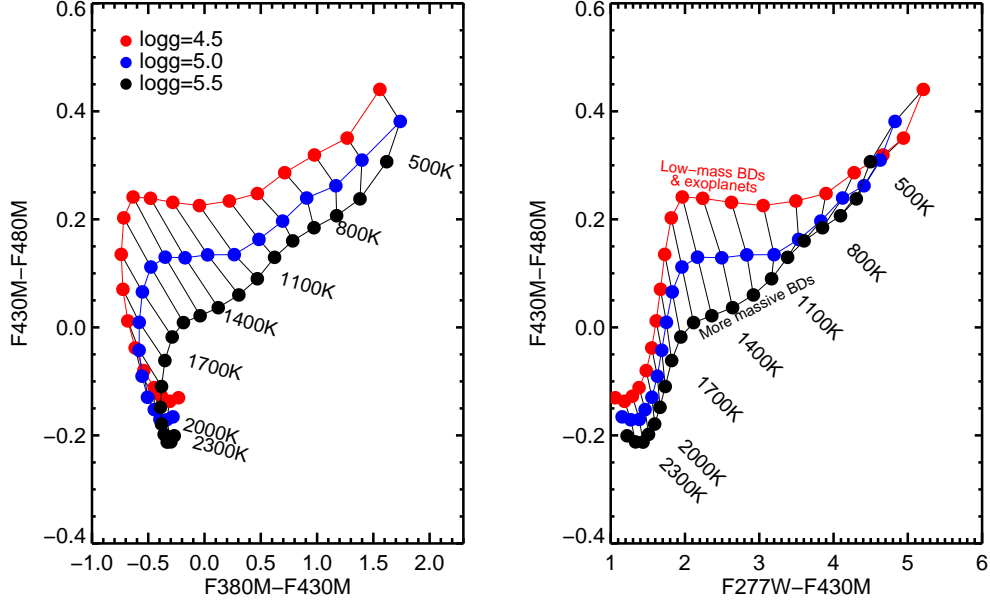}
\end{tabular}
\end{center}
\caption{ \label{fig4} Color-color diagrams for the F380M, F430M and F480M filters (left) and F277W, F430M and F480M filters (right) for brown dwarf and planetary-mass objects of varying effective temperature (500\,K to 2300\,K) and surface gravity ($\log g=4.5$, $5.0$ and $5.5$). The choice of filters for the AMI mode is such that the position of an object in these color-color diagrams constrains its bulk physical properties. }
\end{figure}

The 3 main filters for NIRISS AMI where chosen as, within the domain of interest of this mode ($2.5-5.0\mu$m), they provide the best constraints on the properties of ultracool companions to young stars.  Furthermore, the filters were chosen with relatively narrow bandpasses as to preserve the non-redundancy of $u-v$ coverage. A fourth filter can be used, F277W, that samples a deep water feature centered at 2.7$\mu$m. Figure~\ref{fig3} illustrate the bandpasses of these filters relative to the theoretical spectrum of a 900\,K object at the brown dwarf/planetary mass limit. Within this filter the NIRISS AMI PSF is slightly undersampled and $u-v$ coverage is partially redundant, but it will nevertheless provide useful information at moderate contrasts. The F380M filter samples a relatively clear part of the spectrum, avoiding deep absorption by CO, CH$_4$ and water while F430M and F480M sample a CO feature; this leads to increasingly red $m_{\rm F380M}-m_{\rm F430M}$ and $m_{\rm F380M}-m_{\rm F480M}$ colors for cooler objects. The  $m_{\rm F380M}-m_{\rm F430M}$ color is largely a tracer of effective temperature for objects cooler than $\sim$$1600$\,K (see figure~\ref{fig4}, left panel), objects hotter than $\sim$$1600$\,K have a near-constant $m_{\rm F380M}-m_{\rm F430M}$ color. The depth and shape of the CO and water features probed by the F430M and F480M are both temperature and surface-gravity dependent. For objects cooler than $\sim$$2000$\,K, the  $m_{\rm F430M}-m_{\rm F480M}$ color increases monotonically with decreasing temperature, but low gravity objects ($\log g\sim$$4.5$ versus $5.5$) are systematically redder by $\sim$$0.2$\,mag. For temperatures below $\sim$$1700$\,K, the position of an object in the F380M, F430M and F480M color-color diagram uniquely constrains both its surface gravity (and therefore mass) and temperature.

At higher temperatures ($\sim$$2000$\,K or a spectral type among early Ls), there is degeneracy between temperature and surface gravity that can be lifted by using photometric measurements with the F277W filter (figure~\ref{fig4}, right panel). As it probes an absorption band that becomes largely saturated for ultracool dwarfs, the F277W filter is mostly relevant for hotter objects (T$_{\rm eff}>1000$\,K); for cooler objects, colors combining the F277W and longward filters become very red (e.g., $m_{\rm F277W}-m_{\rm F430M}>4$), and detections at F227W become challenging, leading only to upper limits. As the broad bandpass and coarse sampling of the F277W filter leads to a degraded performance, it is convenient that its principal use will be to determine the properties of the hottest, and therefore brightest, companions.

\section{Saturation limit}

Saturation limits for AMI are important in determining the extent to which it can be used on the brightest targets. The useful limit of the AMI mode is difficult to estimate, as the relevant information on the high spatial frequencies is preserved in the outer parts of the point-spread function and this, in principle, be recovered even if the PSF core is saturated. Image-plane extraction techniques are being explored to take advantage of this fact, but we conservatively assume that the data analysis for the AMI will be done in the Fourier domain, which requires an unsaturated point-spread function core.

Saturation has been estimated by dithering a point-spread function at 25 field positions uniformly using the WebbPSF tool\footnote{http://www.stsci.edu/jwst/software/webbpsf}. We generated PSFs on a grid of 25 spatial positions sampling pixels by 0.2\,pix step, and used 15 wavefront error maps (144, 162 and 180\,nm RMS for the JWST pupil with 5 random phases each) to generate a total of 375 PSFs. For each PSF, we determined the ratio of the brightest pixel to the total flux in the PSF, to establish a saturation magnitude. We assumed a $80\times80$ pixel subarray for the AMI mode with a 41\,ms readout-time ($10\mu$s per pixel) and a saturation at half-well (35\,000\,e$^-$). Only a small fraction of the entire PSF flux falls within the brightest pixel, and this fraction varies depending on the PSF centering relative to the pixel grid. To convert the saturation limit to the WISE photometric bandpasses\footnote{http://wise2.ipac.caltech.edu/docs/release/prelim/expsup/sec4\_3g.html}, we assume a Jeans tail stellar flux distribution ($\lambda^{-4}$ decrease), which is a reasonable assumptions for stars without emission lines. Saturation limits are reported in the WISE-1 for F277W and F380M and WISE-2 for F430M and F480M. The use of the WISE system to establish saturation levels has the advantage that one can readily know whether a given star should saturate as photometric measurements in this system are available for the entire sky. Table~\ref{tbl1} summarizes the properties of the NIRISS AMI filters and their respective saturation limits.

Only for nearby and relatively early-type stars does saturation become an issue. For example, detection of companions around stars in the Upper Scorpius association or the Pleiades (both at $\sim130$\,pc, distance modulus of $\sim5.6$), a saturation limit of $W1=3.7$ (worst case among the nominal filters for AMI) implies an M$_{W1}\sim-1.8$, which corresponds to a late-B star. Planet searches around later-type stars in both regions will not be problematic. Around a late-B star in either group, companions at contrasts of $\sim9$ magnitudes would be of $\sim$M6 spectral type and can be studied from the ground on 8\,m-class telescopes in $L$ and $M$ bands as thermal background is not an issue for such bright targets, so there is, overall, no gap in the $\sim 4\mu$m coverage for these associations.
 
\begin{figure}
\begin{center}
\begin{tabular}{c}
\includegraphics[height=7cm]{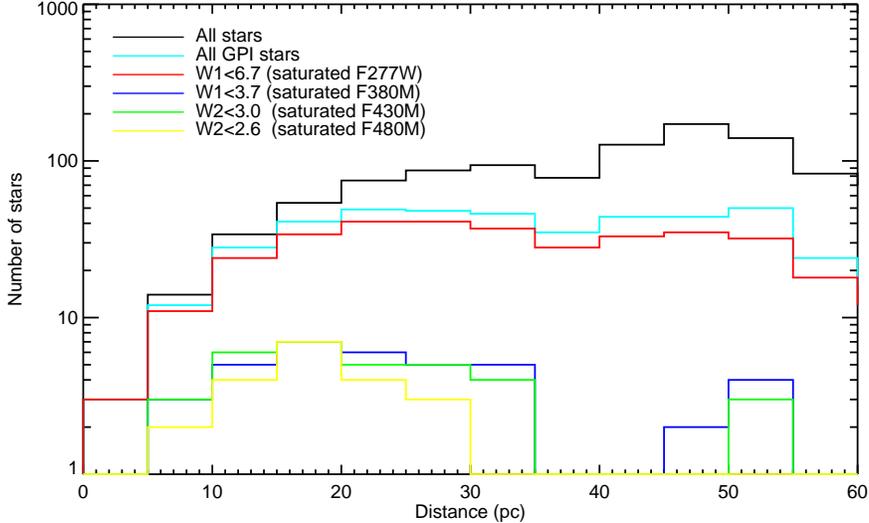}
\end{tabular}
\end{center}
\caption{ \label{fig7} Histogram of distance to young and suspected young stars ($<120$\,Myr) within 60\,pc (black). Cyan histogram shows stars that are amenable to high-contrast imaging with GPI; from this sample, about $50\%$ of all GPI targets will be saturated in F277W (red), but only a few tens saturated in redder filters.}
\end{figure}

For nearby young moving groups, saturation can be an issue for Solar-type stars, for example, at 30\,pc, the saturation limit of $W1=4.2$ corresponds to $M_{\rm W1}=1.8$, or an early-A spectral type. Figure~\ref{fig7} illustrates the sample of young ($<$$120$\,Myr) nearby ($<$$60$\,pc) stars. The sample of 963 stars and brown dwarfs includes both confirmed and suspected young targets of all spectral types. Only a handful early-type young star beyond $\sim35$\,pc are expected to saturate the AMI in the 3 longer filters, and a total of 40, 37 and 24  targets will saturate in F380M, F430M and F480M filters respectively. A much larger fraction of the sample, 337 stars, is expected to saturate in the F277W filter as it is broader, bluer (i.e. closer to the black body peak of stars) and has a coarser pixel sampling (more flux falls within central pixel of the PSF). More than half of the 424 stars observable with GPI (roughly $I\sim9$) will therefore be challenging to observe at 2.77\,$\mu$m and will require image-plane retrieval techniques to detect their faint companions.

\begin{table}[htbp]
\caption{NIRISS NRM filter properties} 
\label{tbl1}
\begin{center}       
\begin{tabular}{|c|c|c|c|c|c|c|} 
\hline
Filter& $\lambda$  & $\Delta\lambda/\lambda$ & IWA$^1$ & OWA$^2$& Fraction of flux &  Saturation \\
      & & & & &in brightest pixel (max)& (35\,000e$^{-}$ ) \\
\hline
\hline
F277W  & 2.78$\mu$m & 26.3\% & 54\,mas & 216\,mas & 0.052 & $W1=7.2$ \\
\hline
F380M  & 3.83$\mu$m & 5.4\% & 74\,mas & 297\,mas & 0.028 & $W1=4.2$ \\
\hline
F430M  & 4.29$\mu$m & 5.0\% & 84\,mas & 336\,mas & 0.022 & $W2=3.5$ \\
\hline
F480M  & 4.82$\mu$m & 6.4\% & 94\,mas & 375\,mas & 0.018 & $W2=3.1$ \\
\hline

\end{tabular}
\\
$^1$Inner working angle as defined $\lambda/(2D)$, where $D$ is the longest AMI baseline ($5.28$\,m). Structures slightly smaller than this scale can be measured under very simple geometry, but sensitivity decreases rapidly (see figure~\ref{fig5}).\\
$^2$Outer working angle where structures become resolved by individual NRM mask holes and pupil ceases to be non-redundant.

\end{center}
\end{table}

\section{NIRISS planet detection limits}

\begin{figure}[!ht]
\begin{center}
\begin{tabular}{c}
\includegraphics[height=9cm]{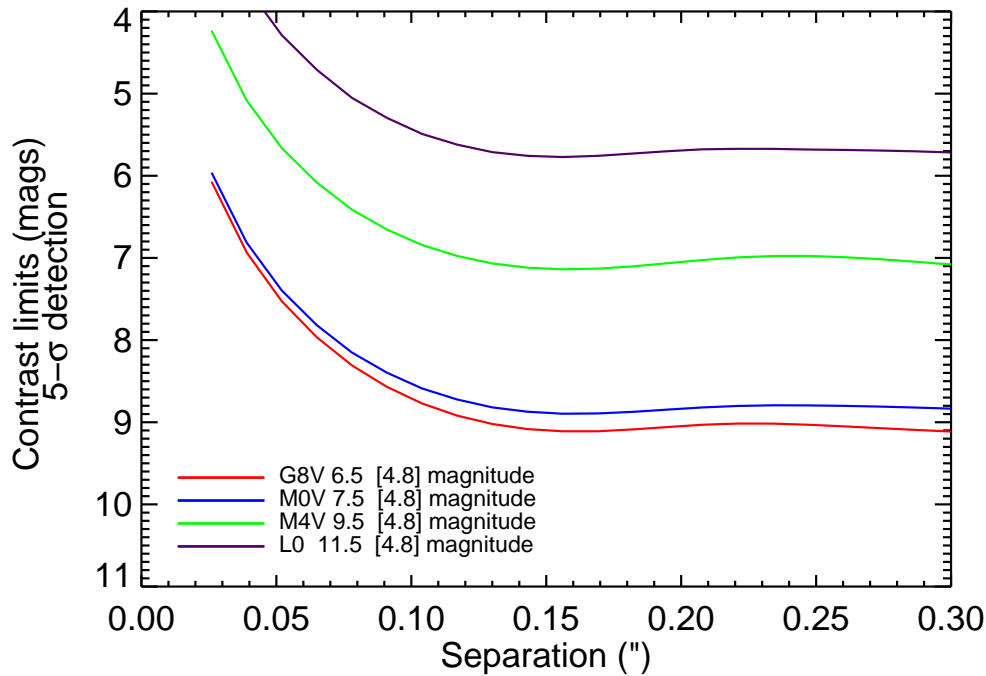}
\end{tabular}
\end{center}
\caption{ \label{fig5} Contrast curves for host stars of various brightnesses ($m_{\rm F480M}=6.5$, $7.5$, $9.5$, $11.5$). For a host distance of 30\,pc and an age of 120\,Myr, typical of many young stars in our neighborhood, this would correspond to host spectral types of G8V, M0V, M4V and L0. The saturation of the gain in contrast for bright targets shows that we are reaching a regime around $m_{\rm F480M}=6.5$ where photon-noise is no longer the dominant cause of error; contributions such as intrapixel response and flat-field accuracy then become important.}
\end{figure}

We performed a detailed simulation of the NRM AMI mode to determine the sensitivity of this mode. The simulations were performed using the numerical recipes described by Sivaramakrishnan et al.\cite{Sivaramakrishnan:2012} assuming a 1\% bad pixel fraction, a $5\times10^{-3}e^-$/s dark current and an on-target integration time of 1\,h. Model-fitting of the scene was performed in the pupil plane. The $5$-$\sigma$ contrast curves shown in figure~\ref{fig5} illustrate the sensitivities for a range of stellar brightnesses. Assuming a distance of 30\,pc and an age of 120\,Myr, typical of many targets in the nearby young associations (e.g. Malo et al.\cite{Malo:2013} and references therein), contrasts of $\sim$$9$\,mags around Sun-like stars are possible, reaching planetary-mass companions with an absolute magnitude of M$_{\rm {4.8}}\sim$$13$ in the Vega system. In a young association, an M$_{\rm {4.8}}\sim$$13$ companion would have $\sim$T7 spectral type\cite{Dupuy2012} and mass about $6$ times that of Jupiter according to substellar evolution models\cite{Chabrier:2000}, not unlike the free-floating planetary-mass member of AB Doradus member CFBDSIR2149\cite{Delorme:2012}. The contrast for cooler hosts, stellar or substellar, leads to the detection of even cooler companions. Around a mid-M primary, the 7 magnitude contrast allows the detection of companions down to M$_{\rm {4.8}}\sim$$14$, which corresponds to $\sim$T9 spectral type and masses of 3-4\,M$_{\rm Jup}$, lighter than any directly detected planetary companion. Around substellar companions, tight binaries comparable to 2M1207AB\cite{Chauvin:2005a} can be detected with companions as light as $1-2$\,M$_{\rm Jup}$. This is especially interesting in the view of the discovery of numerous very low mass member of nearby moving groups\cite{Gagne:2013c, Gagne:2014a}, and the recent discovery of an object at the planet/brown dwarf binary around one of these members\cite{Delorme:2013}.

\section{NIRISS NRM as a follow-up tool for ground-based extreme AO surveys}
\begin{figure}
\begin{center}
\begin{tabular}{c}
\includegraphics[height=7cm]{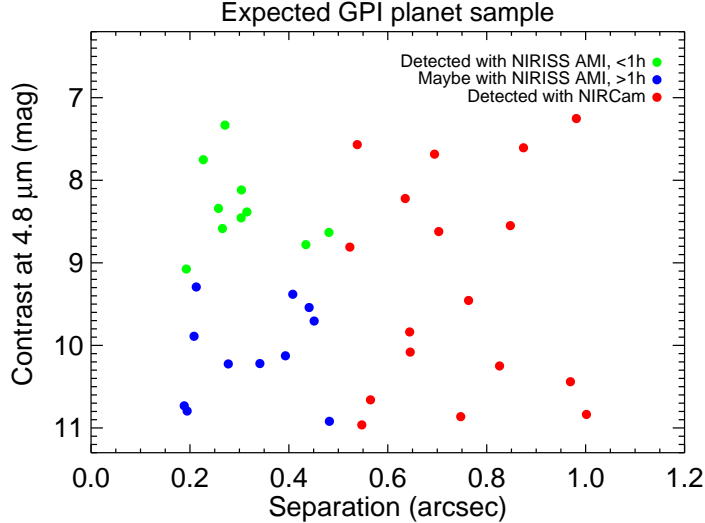}
\end{tabular}
\end{center}
\caption{ \label{fig6} Contrast at 4.8\,$\mu$m versus separation for a typical sample of planets expected from the GPI campaign. Planets beyond $0.5^{\prime\prime}$ will be detectable with NIRCam coronagraph while NIRISS AMI provides complementary spatial coverage between $0.1$ and $0.5^{\prime\prime}$ for the brightest planets.}
\end{figure}

The GPI\cite{Macintosh:2014} (Gemini) and SPHERE\cite{Beuzit:2008, Beuzit:2010} (VLT-ESO) are being used to undertake large-scale surveys for self-luminous exoplanets around nearby young stars. This consists of a sizeable investment in observing time, 900\,h in the case of GPI and 500 nights for SPHERE. Low-resolution spectroscopy of these planets will be obtained in the near-infrared ($1-2.5\mu$m), but mid-infrared observations will be very challenging (e.g., $L$-band observations of HR8799's planets\cite{Skemer:2012, Marois:2010, Galicher:2011}) or impossible for ground-based observatories due to their high thermal background. The $2.5-5.0\mu$m domain is key in establishing the effective temperature of exoplanets expected to be uncovered (typically $400-1000$\,K), as their spectral energy distribution (SED) peaks at those wavelengths. Photometry within this domain is also sensitive to metallicity enhancements that are expected from planets forming through core accretion. Furthermore, brown dwarfs within this domain are known to show out-of-equilibrium carbon chemistry (traced by CO versus CH$_4$ ratio), leading to diagnostic changes in the $\sim$$4\mu$m SED (see in particular Figure 6 in Skemer et al.\cite{Skemer:2012}). Simulations of the expected yield of the GPI exoplanets have been performed and transposed to the NIRISS AMI filters. GPI will be sensitive to planets at projected separations of $0.15$ to $\sim$$1^{\prime\prime}$, the inner half of this domain being inside the AMI's outer working angle. Planets beyond $\sim$$0.5^{\prime\prime}$ will be detectable over the same wavelength domain through NIRCam's coronagraphic mode, providing a complementary coverage. As shown in figure~\ref{fig6}, roughly half of the planets with separations smaller than $0.5^{\prime\prime}$ will be detected in $<1$\,h with the NIRISS AMI. Planets at larger contrasts ($9-11$ mag) may be detectable using longer sequences, but establishing whether this is realistic or not when taking into account systematics remains to be established.


\section{GAIA follow-ups}
\begin{figure}
\begin{center}
\begin{tabular}{c}
\includegraphics[height=6.5cm]{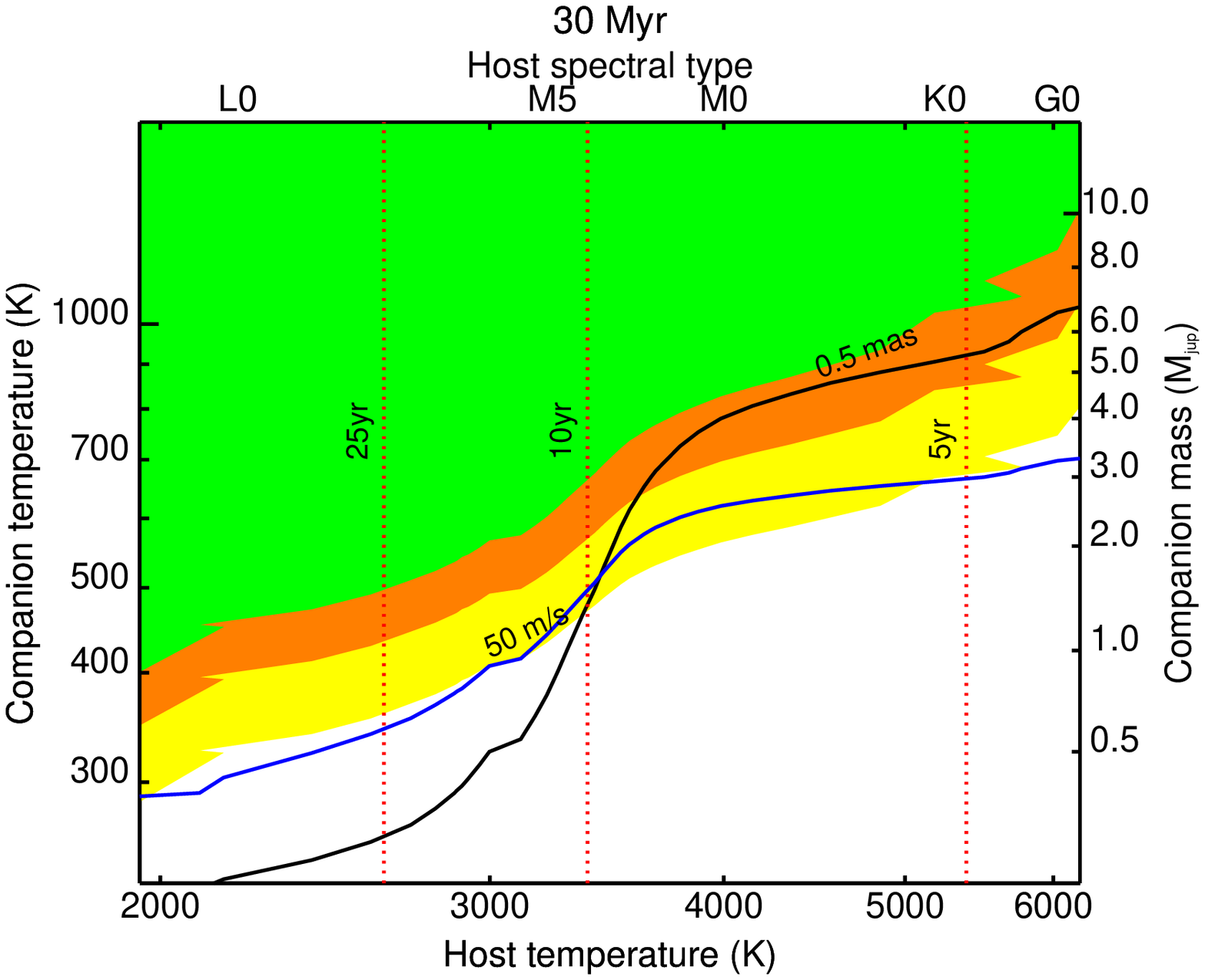}
\includegraphics[height=6.5cm]{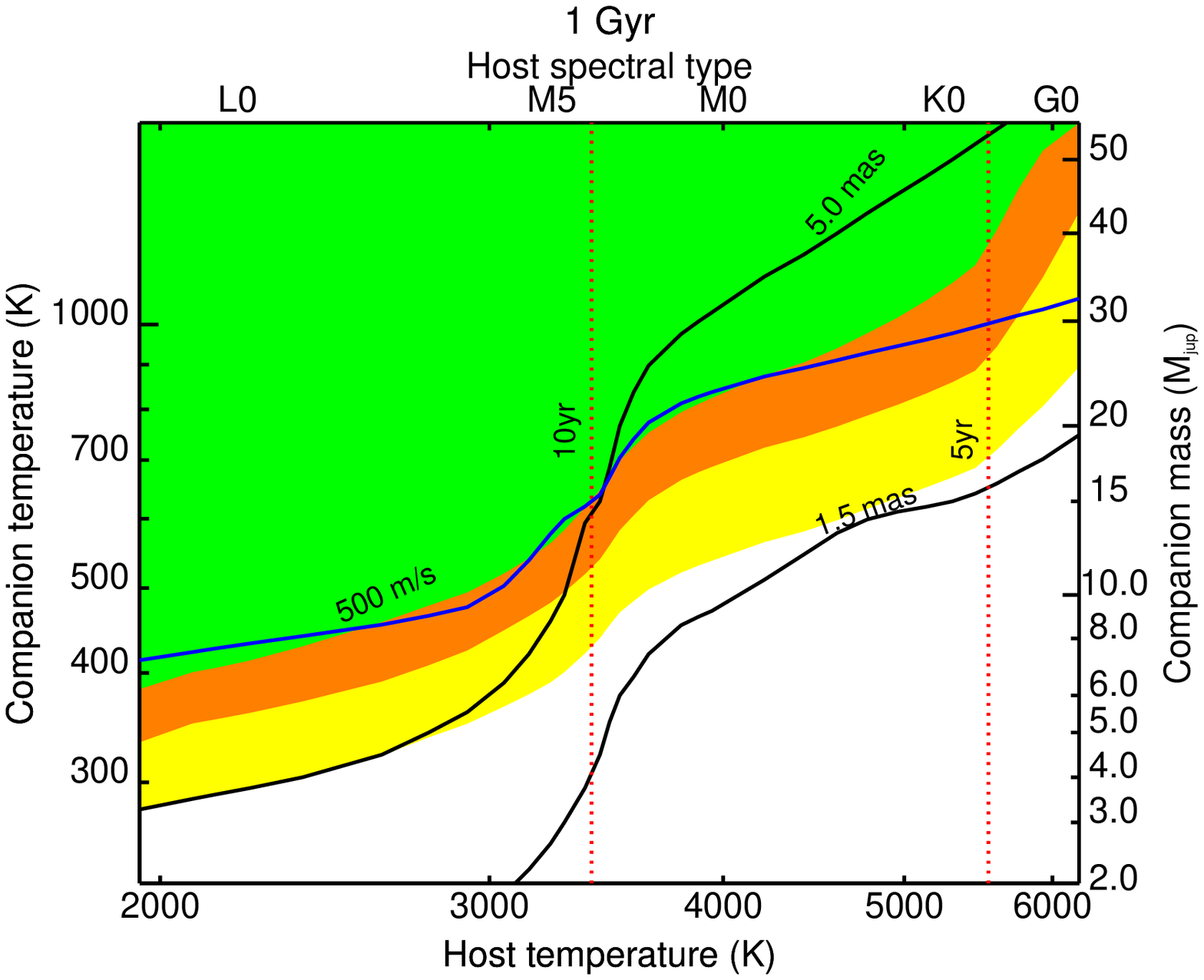}
\end{tabular}
\end{center}
\caption{ \label{fig8} Detection limit for the NIRISS AMI mode expressed as a function of host temperature and spectral type versus companion mass. Here we assume that stars are located at 30\,pc, a distance typical of many young nearby stars and that the companion has a separation of $0.1^{\prime\prime}$, corresponding roughly to the separation at which NIRISS AMI reaches a plateau in sensitivity (see figure~\ref{fig5}). Green, orange and yellow areas correspond to a signal-to-noise ratio of $>10$, $5-10$ and $3-5$ respectively. Dashed lines indicate the orbital period at that separation, ranging from a few years for the earliest-type hosts to $>25$ years for L-type hosts. Astrometric and typical radial-velocity signal for targets close to the detection limit are also shown. The left panel shows the case for a 30\,Myr-old system while the right panel shows the case of a Gyr-old system. Note that the mass scale on the two graphs differ, companions at a given temperature are much more massive at later ages as low-mass companions cooldown on timescales of a few hundred million years.  }
\end{figure}

The GAIA\footnote{http://sci.esa.int/gaia/} astrometric mission successfully launched by ESA  in December 2013. GAIA will provide exquisite astrometry for 1 billion stars and uncover\cite{Casertano:2008} about 4000 planets within 100\,pc in the $1.5-13$\,M$_{\rm Jup}$ mass range. Some of these stars will be young ($<120$\,Myr), and their planetary companions could be detected with the AMI. These planets occupy a unique parameter space: their true (i.e. not model-dependent) masses can be determined directly from the GAIA dataset alone or in combination with radial-velocity follow-ups, and they can be directly detected with the NIRISS AMI. Here, angular resolution is key, planets at $0.1^{\prime\prime}$ around a mid-K star 30\,pc from the Sun orbit their host in $\sim$$5$\,yr (see figure~\ref{fig8}) and are detectable with the NIRISS AMI. To be detected with NIRCam's coronagraph\cite{Green:2005, Beichman:2010a}, the prime instrument aboard JWST to detect more distant companions at these wavelengths, the planet would have to be at $0.5-1.0^{\prime\prime}$, depending on contrast thus leading to orbital periods of 11 to 31 times longer. NIRISS AMI therefore has a unique niche, in being able to detect planets that can have their dynamical mass measured in a manageable time frame ($< 25$\,yr). As shown in figure\,\ref{fig8} for stars in 30\,Myr-old  associations (Tucana-Horologium, Columba \& Carina), the NIRISS AMI will be able to detect planets as light as $1-3$\,M$_{\rm Jup}$ depending on host brightness. These planets induce reflex motions of $>0.5$\,milliarcsecond (defined as the semi-major axis of the planet's motion), amply above the GAIA detection threshold. These planets also induce line-of-sight radial-velocity changes of $>50$\,m/s, readily detectable by stabilized RV spectrographs. Planets as light as $\sim$$4$\,M$_{\rm Jup}$ can be found around Gyr-old L dwarfs (see figure~\ref{fig8}, right), while brown-dwarf mass $>13$\,M$_{\rm Jup}$ companions will be found around hosts earlier than mid-M.

\section{AGN science with NIRISS}

\begin{figure}
\begin{center}
\begin{tabular}{c}
\includegraphics[width=16cm]{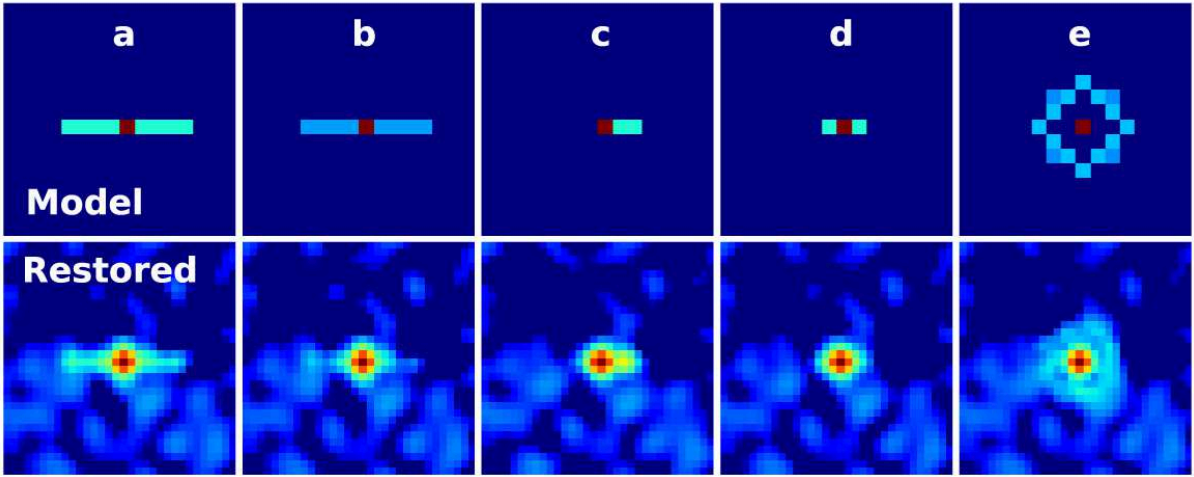}
\end{tabular}
\end{center}
\caption{ \label{fig9} Synthesis imaging of Active Galactic Nuclei (AGN) environs using NIRISS AMI (after Ford et al. 2014). Simple AGN models consisting of a point-source nucleus and fainter extended structure thin bars and rings are shown in the top row, on the NIRISS 65-mas pixel scale. At a distance of 10 Mpc one detector pixel spans 3.2 pc. The corresponding restored images (bottom row) combine two observations oriented perpendicularly. Such orientations would be spaced by about a quarter of the Webb orbital period. The restoration process utilizes a finer pixel scale than the detector pixels. The Multichannel Image Reconstruction Image Analysis and Display (MIRIAD) software package CLEAN algorithm was used for these image reconstructions. This simulation suggests that a 100:1 pixel-to- pixel contrast is well within the reach of NIRISS AMI with expected levels of noise. Explanations: (a) The integrated flux from the bar is 1 magnitude fainter than the point source. (b) The bar is 2 magnitudes fainter than the point source. (c) An asymmetric 3-pixel bar with the same surface brightness as the bar in (a), and (d) a similar symmetric bar. (e) A 5-pixel diameter circular ring with integrated flux 1 magnitude fainter than the point source.}
\end{figure}

Accretion onto super-massive black holes (SMBHs; $10^6 -10^9$ M$_\odot$) at the centers of active galactic nuclei (AGNs) is believed to play a key role in Lambda-Cold Dark Matter ($\Lambda$CDM) cosmology and galaxy formation\cite{Silk:2010}. However, the extreme luminosities of AGNs\cite{Kauffmann:2009} ($10^{10}-10^{16}$ L$_{\odot}$) make imaging of their surrounding a challenge. The problem is exacerbated by the small angular size of the parsec-scale accretion flows at typical AGN distances, often more than 10\,Mpc. Ford et al. (2014)\cite{Ford:2014} demonstrate that true imaging at $4.5\mu$m will enable NIRISS AMI to study the environments of accreting SMBHs. After including realistic NIRISS and Webb noise sources and operational plans, Ford et al. report that AMI will map extended structure at 70\,mas angular resolution and a dynamic range of 100:1 (see figure~\ref{fig9}).  Such data can test models of AGN binarity, feedback, fueling and structure, and will complement longer wavelength observations of, for example, the Atacama Large Millimeter/submillimeter Array.  Comparable high-resolution imaging of AGN is not possible with any ground-based infrared interferometer or telescope for at least three reasons: (a) thermal background on ground-based telescopes limits them to a handful of bright targets, (b) fluctuating amplitude and phase error due to atmospheric variability hamper credible image reconstruction unless a prior model of the object is invoked, and (c) good adaptive-optics compensation is limited to the brightest few targets. NIRISS AMI can complete a survey of a dozen bright AGNs with a total of about 200 minutes of exposure, in spite of the 15\% throughput of the non-redundant mask.  The cosmological model of hierarchical galactic mergers predicts an abundance of SMBH mergers. If merging black holes continue to accrete, then binary AGN in galaxies should be commonplace. Nevertheless, very few such nuclei have been observed to date \cite{Liu:2010a}. NIRISS AMI will resolve binary AGN as close as 70 mas or 17 pc at a distance of 50 Mpc. This performance improves upon current ground-based optical or IR interferometry, especially when dust obscures the AGN in the shorter $JHK$ near-infrared bandpasses.

\acknowledgments        
 
This work is financially supported in part by the Canadian Space Agency (contract 9F052-130055/001/MTB), by NASA grant APRA08-0117 and the National Science Foundation Graduate Research Fellowship Program under Grant No. DGE-1232825.

\bibliography{bibdesk}
\bibliographystyle{spiebib}  

\end{document}